\documentclass[12pt]{article}
\usepackage{latexsym}
\usepackage{amssymb,amsmath}
\usepackage{graphicx}
\usepackage{color}
\usepackage{hyperref}

\newcommand{\be}{\begin{equation}}
\newcommand{\ee}{\end{equation}}

\def\la{\mathrel{\mathpalette\fun <}}
\def\ga{\mathrel{\mathpalette\fun >}}
\def\fun#1#2{\lower3.6pt\vbox{\baselineskip0pt\lineskip.9pt
  \ialign{$\mathsurround=0pt#1\hfil##\hfil$\crcr#2\crcr\sim\crcr}}}

\newcommand{\etal}{{\em et al.\,}}                                              
\newcommand{\etalref}{{\em et al.}}

\newcommand{\etc}{{\em etc.}}                                                   
\newcommand{\eg}{{\em e.g.,\,}}

\newcommand{\beq}{\begin{equation}}                                             
\newcommand{\eeq}{\end{equation}}     
\newcommand{\epm}{\ensuremath{e^{\pm}}}
                                          
\newcommand{\sigv}{\ensuremath{\langle \sigma v\rangle}}

\newcommand{\omm}{\ensuremath{\Omega_{\rm M}}}
\newcommand{\omdm}{\ensuremath{\Omega_{\rm DM}}}
\newcommand{\omb}{\ensuremath{\Omega_{\rm B}}}
\newcommand{\omr}{\ensuremath{\Omega_{\rm R}}}
\newcommand{\omk}{\ensuremath{\Omega_{\rm k}}}
\newcommand{\oml}{\ensuremath{\Omega_{\Lambda}}}

\newcommand{\omhhb}{\ensuremath{\Omega_{\rm B}{h}^{2}}}

\newcommand{\etab}{\ensuremath{\eta_{\rm B}}}
\def\3he{$^3$He}
\def\4he{$^4$He}
\def\6li{$^6$Li}
\def\7li{$^7$Li}
\def\Yp{Y$_{\rm P}$}

\title{Is The Universal Matter - Antimatter Asymmetry Fine Tuned?}
\author{Gary Steigman and
Robert J. Scherrer\footnote{Following the
untimely death of Gary Steigman, the second author was brought in to complete this chapter.
He has endeavored to adhere as closely as possible to the original format and spirit of the manuscript
constructed by
the first author.}}

\begin{document}
\maketitle

\begin{abstract}

The asymmetry between matter and antimatter (baryons and antibaryons or nucleons and antinucleons, along with their
accompanying electrons and positrons) is key to the existence and nature of our Universe.  
A measure of the matter - antimatter asymmetry of the Universe is provided by the present value of the universal ratio
of baryons (baryons minus antibaryons) to photons (or, the ratio of baryons to entropy).  The baryon asymmetry parameter
is an important physical and cosmological parameter.  But how fine tuned is it?  A ``natural" value for this parameter is
zero, corresponding to equal amounts of matter and antimatter.  Such a Universe would look nothing like ours and would
be unlikely to host stars, planets, or life.  Another, also possibly natural, choice for this dimensionless parameter
would be of order unity, corresponding to nearly equal amounts (by number) of matter (and essentially no antimatter) and
photons in every comoving volume.  However, observations suggest that in the Universe we inhabit the value of this
parameter is nonzero, but smaller than this natural value by some nine to ten orders of magnitude.  In this
contribution we review the evidence, observational as well as theoretical, that our Universe does {\it not}
contain equal amounts of matter and antimatter.  An overview is provided of some of the
theoretical proposals for extending the standard models of particle physics
and cosmology in order to generate such an asymmetry during the early evolution of the Universe.

Any change in the magnitude of the baryon asymmetry parameter necessarily leads
to a universe with physical characteristics different from those
in our own.  Small changes in this parameter will barely affect cosmic evolution, while large changes might
alter the formation of stars and planets and affect the development of life.  The degree of fine tuning in the baryon asymmetry parameter
is determined by the width of the range over which it can be varied and still allow for the existence of life.
Our results suggest that the baryon asymmetry parameter can be varied over a
very wide range without impacting the prospects for life; this result is {\it
not}
suggestive of fine tuning.

We note that according to those extensions of the standard models of particle
physics and cosmology that allow for a nonzero baryon number, the Universe began
with zero baryon number, at a time (temperature) when baryon number was
conserved.  As the Universe expanded and cooled, baryon number conservation was
broken at some high temperature (mass/energy) scale and a nonzero baryon number
was created.  However, even though baryon nonconservation is strongly suppressed
at late times (low temperatures), baryon number is not conserved, so matter
(protons, the lightest baryons) might eventually decay, with the baryon number
reverting back to zero.  Ashes to ashes, dust to dust, the Universe began with
zero baryon number and may well end that way.
 
\end{abstract}

\section{Introduction and Overview}
\label{sec:intro}

The asymmetry between matter and antimatter (baryons and antibaryons or nucleons and antinucleons, along with their
accompanying electrons and positrons) is key to the existence and nature of our Universe.  Any causal Lorentz-invariant
quantum theory allows for particles to come in particle-antiparticle pairs.  The discovery of the antiproton
\cite{chamberlain} in 1955 quickly stimulated serious consideration of the antimatter content of the Universe
\cite{goldhaber,alpherherman} and led to  constraints on the amount of antimatter based on the astrophysical effects of
interacting matter and antimatter \cite{burbidgehoyle}.  At the time, and for many years after, the prevailing view in
the physics community was that baryon number (the quantum number that distinguishes baryons and antibaryons)
was absolutely conserved, and this assumption led to two differing points of
view.  Either the Universe is and always has been symmetric between matter and antimatter, or the Universe is
and always has been asymmetric, with an excess of matter over antimatter that has remained unchanged from the beginning
of the expanding Universe (the big bang).  Those who believed the Universe to be symmetric between matter and antimatter
were undeterred by the fact that that the only antimatter seen up to that time (not counting positrons) was the handful
of antiprotons created in collisions at high energy accelerators.  Those who believed the Universe to be asymmetric had
to come to grips with the dilemma of creating such a Universe if the laws of physics dictated that particles are always
created (and destroyed) in pairs and that baryon number is absolutely conserved.

Most ignored this dilemma.  Andrei
Sakharov \cite{sakharov} did not.  To set the stage for Sakharov's seminal work, it is useful to recall the 1965
discovery of the cosmic microwave background (CMB) radiation \cite{penzwil,dicke}, which transformed the study of
cosmology from philosophy and mathematics to physics and astronomy.  It quickly became clear that the discovery of the
radiation content of the Universe, along with its observed expansion, ensured that very early in its evolution, when the
temperature and densities (both number and energy densities) were very high, collisions among particles would be very rapid and energetic and, at sufficiently high temperatures, particle-antiparticle pairs would be produced (and would annihilate).  Sakharov explored the requirements necessary for such high energy collisions in the early Universe to create a matter-antimatter asymmetry if none existed initially.  Sakharov's recipe for cooking a universal baryon asymmetry has three ingredients.  One obvious condition is that baryon number cannot be absolutely conserved; baryon number (B) conservation must be violated.  Although the standard model of particle physics at the time did not allow for violation of baryon number conservation, the later development of grand unified theories (GUTs) did.  Sakharov also noted that the discrete symmetries of parity (P) and charge conjugation (C), replacing particles with antiparticles, or of CP, would need to be broken as well.  Current models, in agreement with accelerator data, do allow for P and CP violation.  Sakharov's third ingredient is not from particle physics, but from cosmology, relying on the expansion of the Universe.  The third ingredient in the recipe requires that thermodynamic equilibrium not be maintained when the B, P, and CP violating collisions occur in the early Universe.  Although at the time of Sakharov's work there was no evidence that conservation of B and CP were violated, it was already known that parity is not conserved in the weak interactions and that the expansion of the Universe could possibly provide the required departure from thermodynamic equilibrium.
Sakharov set the stage for consideration of a Universe with unequal amounts of matter and antimatter.
We will revisit Sakharov's three conditions for baryogenesis in \S \ref{sec:sakharovetal}.

In the hot, dense thermal soup of the very early Universe, matter and antimatter (baryons and antibaryons) are as
abundant as all the other particles whose mass is less than the temperature.  As the Universe expands and cools,
particle-antiparticle pairs annihilate, leaving behind only the lightest particles, along with any particle-antiparticle
pairs that evaded annihilation in the early Universe or, perhaps, in an asymmetric Universe, an initial matter excess
that escaped annihilation.  In the late Universe, when the temperature (in energy units) is far below the masses of the
unstable particles of the standard model (SM) of particle physics, only photons and the lightest stable (or very long
lived) SM particles remain: nucleons (and possibly antinucleons), electrons (and possibly positrons), and the three SM
neutrinos.  In cosmology it is conventional to refer to all ordinary matter consisting of
nucleons and electrons (nuclei, atoms, and molecules), as
``baryons" (B)\footnote{Throughout this article the terms baryons, nucleons, ordinary matter, and normal matter are used
interchangeably.} to distinguish it from dark matter (DM).  Electrons are not baryons, but
their (very small) contribution to the present-day matter density is included in this definition of the baryon density.
The photons and neutrinos are often
referred to as ``radiation".  The matter-antimatter asymmetry is the difference between the numbers of baryons and
antibaryons.  Since this is an extensive quantity, scaling with the size of the volume considered, it is useful to
introduce the ratio (by number) of baryons to photons to quantify the size of any matter-antimatter asymmetry.  The
ratio of the baryon (minus the antibaryon) and photon number densities, $\etab = n_{\rm B}/n_{\gamma}$, provides a
measure of the matter-antimatter asymmetry of the Universe.  However, as the Universe expands and cools, the heavier,
unstable SM particles annihilate and decay, increasing the number of photons $N_\gamma$ in a comoving volume $V$, where
$N_{\gamma} = n_{\gamma}V$, while the baryon number in the same comoving volume is unchanged (at least during those
epochs when baryons are conserved).  Instead, it is the entropy, $S = sV$, in the comoving volume, not the number of
photons, that is conserved as the Universe expands adiabatically.  The entropy and the number of photons in a comoving
volume are related by $S =1.8g_{s}N_{\gamma}$, where the total entropy is related to the entropy in photons alone by $S
\equiv (g_{s}/2)S_{\gamma}$, and $S_{\gamma} = 4/3(\rho_{\gamma}/T)V = 4/3(\langle E_{\gamma}\rangle/T)N_{\gamma}$ and
$\langle E_{\gamma}\rangle = 2.7\,T$.  The quantity $g_{s} = g_{s}(T)$ counts the number of degrees of freedom
contributing to the entropy at temperature $T$.  For the SM of particle physics, with three families of quarks and
leptons, at temperatures above the mass of the heaviest SM particle (the top quark), $g_{s} \approx 427/4$.  For
temperatures below the electron mass, after the three flavors of weakly interacting neutrinos have decoupled and the
photons have been heated relative to the neutrinos by the annihilation of the \epm~pairs, $g_{s} \rightarrow g_{s0}
\approx 43/11$.  As a result, as the Universe cools from above the top quark mass to below the electron mass, the number
of photons in a comoving volume increases by a factor of $\approx 27$, and the baryon to photon ratio is diluted by this
same factor. In an adiabatically expanding Universe (as ours is assumed to be) the entropy in a comoving volume is
conserved, along with the net number of baryons minus antibaryons (during those epochs when baryon number
nonconservation is strongly suppressed).  Therefore, the ratio of baryon number to entropy, $N_{\rm B}/S = n_{\rm B}/s$,
provides a measure of the baryon asymmetry whose value is unchanged as the Universe expands and cools.  Evaluated in the
late Universe, after \epm~annihilation is complete, $s/n_{\gamma} \rightarrow (s/n_{\gamma})_{0} = 1.8\,g_{s0} \approx
7.0$, so that $n_{\rm B}/s \approx \etab/7.0$.  Consistent with most of the published literature, \etab~is evaluated
here in the late Universe, so that $\etab \equiv \eta_{\rm B0} \equiv (n_{\rm B}/n_{\gamma})_{0}$.  In the discussion
here $\eta_B$ and $n_{\rm B}/s$ will both be referred to as the ``baryon asymmetry
parameter."  

In a matter-antimatter {\it symmetric} Universe the baryon asymmetry parameter $\etab = 0$.  For a
quantity that could, in principle, have any value between $-\infty$ and $+\infty$,\footnote{In a Universe with more ``matter" than ``antimatter", $\etab > 0$.
For the opposite case, where $\etab < 0$, the definitions of matter and antimatter could be interchanged.
Therefore, without loss of generality, it is assumed here that $\etab \geq 0$.}
zero might seem to be a ``natural" choice.  

When is a physical parameter, such as the baryon asymmetry parameter, considered to be fine tuned?  The criteria for
answering this question, along with a discussion of the degeneracies with other physical parameters, are discussed in
\S\ref{sec:groundrules}.  In \S\ref{sec:symmetric} the overwhelming observational and theoretical evidence that our Universe is {\it not} matter-antimatter symmetric is reviewed, excluding the natural choice of $\etab = 0$.  Faced with the necessity that a Universe hosting stars, planets, life requires $\etab \neq 0$, \S\ref{sec:sakharovetal} provides an overview of the multitude of particle physics (and cosmology) models proposed to generate a nonzero baryon asymmetry during the early evolution of the Universe.  These models are capable of generating a baryon asymmetry that is much smaller or much larger than that observed in our Universe, suggesting that there might be universes with almost any nonzero values of \etab.  In an asymmetric Universe the quantitative value of the baryon asymmetry parameter plays an important role in primordial nucleosynthesis (big bang nucleosynthesis: BBN), regulating the abundances of the nuclides produced in the early Universe, before any stellar processing.  BBN is reviewed for a large range of \etab~in \S\ref{sec:bbn}.  The degeneracy of the baryon asymmetry parameter with other cosmological parameters is discussed in \S\ref{sec:cosmo} and a variety of alternate cosmological models allowing for a
range of \etab~values are presented in \S\ref{sec:alternate}.  
The criterion used here
to judge the viability of alternate cosmological models is whether their universes are capable of hosting stars,
planets, and life.
Our results and conclusions  are summarized in \S\ref{sec:conclusions}.

\section{Definition of Fine-Tuning of the Baryon Asymmetry Parameter}
\label{sec:groundrules}

How fine-tuned is the baryon asymmetry parameter?  Here we will adopt a definition of fine tuning based on the capability of
the Universe to harbor life.  Clearly, small changes in the asymmetry parameter will have little effect on cosmic evolution.
However, large changes in this parameter will have major effects, notably altering the production of elements in the early universe and changing
the process of structure formation through the growth of primordial density perturbations.  We will see that the former,
even in extreme cases, is unlikely to have any effect on the development of life in the Universe, while the latter can
have profound effects.  In particular, if the process of galaxy and star formation is too inefficient, then there will be no
planetary systems
to harbor life.  One must be cautious, of course, in defining the limits on environments that can support life; our
argument will be based on life as we observe it, which exists on planets orbiting stars.  It is always possible that more
extreme environments might harbor life in ways that we have not considered; for example, Avi Loeb has pointed out that
the cosmic microwave background can provide an energy source for life when the universe was only 10 million years old
and the temperature of the CMB was between the freezing and boiling points of water \cite{Loeb}.  While we will not consider
such extreme possibilities here, caution is always advised when defining the conditions needed for the existence
of life.

The extent to which the value of $\eta_B$ is fine tuned will depend on how widely it can be varied while still allowing for the existence
of life.  The issue of the fine-tuning of $\eta_B$ is not, of course, a true-false question:
the best we can do is to determine an allowed range for $\eta_B$.
The width of this range can then suggest the plausibility (or lack thereof) of
the need for special initial conditions or special values for the underlying fundamental parameters that determine
$\eta_B$.  But the question, ``Is the baryon asymmetry parameter fine tuned?" does not have a yes or no answer.

In considering the variation of one or more physical parameters, a choice must be made:  do we consider the variation of
the baryon asymmetry parameter alone, or do we allow other parameters to vary at the same time?
In the latter case changes in the value of one
parameter may be compensated, at least in part, by changes in other parameters.

As an example, consider the way in which the relation between the baryon to entropy ratio and
the baryon to photon ratio depends on the number of neutrino flavors, as well as on the neutrino
decoupling temperature, which depends in turn on the strength of the weak interactions.
In an alternate Universe where there are $N_{\nu}$ flavors of neutrinos, instead of the SM value of $N_{\nu} = 3$, $g_{s0} = 43/11 + 7(N_{\nu} - 3)/11 = 43/11(1 + 7(N_{\nu} - 3)/43)$ and $g_{\rho0} = 3.36 + 0.454(N_{\nu} - 3) = 3.36(1 + 0.135(N_{\nu} - 3))$.  For these results it has been assumed that when $N_{\nu} \neq 3$, the usual weak interactions are unchanged and all neutrinos decouple when $T_{dec} \gg m_{e}$ (but $T_{dec} \ll m_{\mu}$), so that $(T_{\nu}/T_{\gamma})_{0} \approx 4/11$.  With these caveats, for $N_{\nu} \neq 3$, the late time entropy per photon is $(s/n_{\gamma})_{0} \approx 7.0(1 + 7(N_{\nu} - 3)/43)$ and the relation between \etab~and $n_{\rm B}/s$ is changed,
\beq
\etab = (n_{\rm B}/n_{\gamma})_{0} \approx 7.0(1 + 7(N_{\nu} - 3)/43)\,(n_{\rm B}/s)\,.
\eeq
For example, in an alternate Universe with only one neutrino flavor ($N_{\nu} = 1$), $\etab \approx 4.7(n_{\rm B}/s)$, while in one with eight flavors of neutrinos, ($N_{\nu} = 8$)\,\,\footnote{For $N_{\nu} \leq 8$, QCD is asymptotically free, allowing for quark confinement and bound nuclei \cite{grosswilczek}.}, $\etab \approx 12.8(n_{\rm B}/s)$.

In general, allowing multiple parameters to vary simultaneously will weaken the constraints provided when only one of them
is varied,
an issue that is likely to be an issue with many of the other essays in this volume.
For example, consider the atomic energy scale,
\beq
\epsilon \equiv \mu_{\rm H}c^{2}\alpha^{2} = \bigg({m_{e}m_{p} \over m_{e} + m_{p}}\bigg)c^{2}\bigg({e^{2} \over \hbar c}\bigg)^{2}\,,
\eeq
where $\mu_H$ is the reduced mass of the proton-electron system, and the fine structure constant is $\alpha = e^{2}/\hbar c \approx 1/137$ (when measured at low energies).  For $m_{e}c^{2} \approx 0.51\,{\rm MeV}$ and $m_{p} \approx 0.94\,{\rm GeV}$, $\epsilon \approx m_{e}c^{2}\alpha^{2} \approx 27\,{\rm eV}$.  Since $\epsilon$ is not a dimensionless parameter, perhaps it is the dimensionless parameter $\epsilon/\mu_{\rm H}c^{2} = \alpha^{2} \approx 5.3\times10^{-5}$ that is fundamental.  Suppose that $\alpha$ and $\mu_{\rm H}c^{2}$ are allowed to change, while the atomic energy scale, $\mu_{\rm H}c^{2}\alpha^{2}$, is kept unchanged.  For example, $m_{e}$ and $m_{p}$ might change while $m_{e}/m_{p} \ll 1$ might be (nearly) unchanged.  Atomic energy levels will be largely unchanged while nuclear energies will be changed.  How much freedom is there to change $\alpha$
along with other fundamental parameters (\eg $m_{e}$, $m_{p}$, $m_{e}/m_{p}$), while leaving most of ``ordinary" atomic and nuclear
physics unchanged?  This issue of ``degeneracy" among physical parameters will rear its head in the subsequent discussion of the fine tuning of the baryon asymmetry of the Universe.  
When exploring model universes
with different values of \etab, we will keep all other parameters (\eg $\alpha$, $m_{e}/m_{p}$, $N_{\nu}$, \etc) fixed.  However,
we need to remain aware that the results presented here can be considerably altered if multiple parameters are simultaneously varied.

\section{The Case Against a Symmetric Universe}
\label{sec:symmetric}

Over the years, experiments at ever higher energies have confirmed that particles are created (and annihilated) in pairs and
that in all collisions studied so far, baryon (and lepton) number is conserved.  Perhaps only at the very highest energies,
inaccessible to the current terrestrial accelerators, or in searches for proton decay, will nonconservation of baryon (and
lepton) number be revealed.  However, it is not unreasonable to ask how our present Universe would differ if baryon number were
absolutely conserved.  A complementary approach is to ask what astrophysical observations can tell us about the amount of antimatter
(if any) in gas, stars, galaxies, and clusters of galaxies in the current Universe (\eg \cite{burbidgehoyle}).  These two approaches
are explored here.  The discussion here is based on several earlier papers by the first
author (\eg \cite{gs69,gs71,gs76,gs79,gs08}); the reader is urged to see those papers for details and for many further references.

\subsection{The Observational Evidence Against a Symmetric Universe}
\label{sec:obs}

To paraphrase remarks by the first author in a 1976 review of the status of antimatter in the Universe \cite{gs76}, it is quite easy
to determine if an unknown sample is made of matter or antimatter.  The most rudimentary detector will suffice.  Simply place your
sample in the detector and wait.  If the detector disappears (annihilates), your sample contained antimatter.  Indeed, if you had
handled your sample, you would have already known the answer.  Astrophysical sources have been repeating this experiment over
cosmological times.  The first lunar and Venus probes confirmed that the Moon and Venus are made of matter, not antimatter.  Indeed, the solar wind, sweeping past the planets of the solar system revealed, by the absence of annihilation gamma rays, that the Sun and the planets, and other solar system bodies are all made of what we have come to define as matter.  Were any of the planets made of antimatter, they would be the strongest gamma ray sources in the sky (if they hadn't already annihilated away).  As may be inferred from the discussion below in \S\ref{sec:annihilation}, if there were any antimatter in the material (the pre-solar system gas cloud) that collapsed to form the planets and other solid body objects in the solar system, it would have annihilated away long before the solar system formed.  The same is true for the stars in our Galaxy.  On theoretical grounds, is is highly unlikely that in a Universe some 14 Gyr old, there are any non-negligible amounts of antimatter surviving in our Galaxy.

In a typical nucleon - antinucleon annihilation, $\sim 5 - 6$ pions are produced.  The pions decay to muons, neutrinos, and
photons and the muons decay to electrons (\epm~pairs) and neutrinos.  The \epm~pairs may annihilate in flight or, being tied to
local magnetic fields, they may lose energy by Compton emission and annihilate nearly at rest (producing a characteristic 511
keV line) \cite{burbidgehoyle}.  Photons from matter-antimatter annihilations provide the most sensitive, albeit indirect, probe
of the presence of antimatter, mixed with ordinary matter, on galactic and extragalactic scales.  In the Galaxy, gas (clouds of
atomic or molecular gas) and stars are inevitably mixed.  If either contained significant amounts of antimatter, the
result would be annihilation,
along with the corresponding production of gamma rays.  The lifetime against annihilation of an antiparticle (\eg an antiproton) in the gas
in the interstellar medium (ISM) of the Galaxy is very short, $t_{ann} \approx 300\,{\rm yr}$ \cite{gs76}.  It is therefore not
surprising that observations of galactic gamma ray emission set very strong constraints on the antimatter fraction in
the ISM, $f_{\rm ISM} \la 10^{-15}$ \cite{gs76}.  There can be no significant amounts of antimatter in the gas in the Galaxy.
 
What about antistars?  When gas collapses to form stars, the annihilation rate grows as the number density while the collapse
rate increases only as the square root of the density.  As a result, unless there were no normal matter in the gas that might
collapse to form an antistar, the antistar would never form.  Setting this aside, let us suppose that antistars had
somehow formed in the
Galaxy.  As the gas in the ISM flowed past these antistars, there would be annihilation, resulting in gamma rays.  Using
by now outdated (40 year old!) gamma ray data, the first author \cite{gs76} determined that the absence of gamma rays indicated that
the nearest antistar in the Galaxy is at least 30 pc away.  This result sets an upper limit on the total number of antistars, $N$, that could
be in the Galaxy: $N < 10^7$, a small fraction of all the stars in the Galaxy.  Although more recent gamma ray data can refine these bounds, the old data were already sufficiently strong to argue against any significant amounts of antimatter in the Galaxy.

Galactic cosmic rays, coming to us from outside of the solar system, provide a valuable direct probe of antimatter in the
Galaxy.  Whatever the sources of the galactic cosmic rays, the discovery of antinuclei in the cosmic rays would provide direct
evidence (a ``smoking gun") for the presence of antimatter in the Galaxy (for more details, but obsolete data, see the
discussion in \cite{gs76}).  The antiproton would be the lightest antinucleus, but in high energy collisions between cosmic rays
and interstellar gas, some ``secondary" antiprotons will be produced.  Indeed, antiprotons have been observed in the cosmic
rays, but their numbers are consistent with a secondary origin.  However, production of more complex antinuclei in high energy
cosmic ray - interstellar gas collisions (secondary antinuclei) is strongly suppressed and, to date, no antideuterons
\cite{antideuteron} or antialpha \cite{ams} particles have been detected in the cosmic rays.  For example, the 1999 AMS upper bound
\cite{ams} to the cosmic ray antihelium to helium ratio is $< 10^{-6}$, providing a strong supplement to the gamma ray data
suggesting our Galaxy has no significant amounts of antimatter.  The absence of primary antinuclei in the cosmic rays is evidence
that the sources of the galactic cosmic rays contain little, if any, antimatter (indeed,
if there were some antimatter mixed with a predominant amount of ordinary matter in the cosmic ray sources, they likely
would have annihilated over the lifetimes of the sources).

What of external galaxies or extragalactic high luminosity sources such as AGNs or QSOs?  If annihilations deposit their energy
locally, then the gamma ray flux and the luminosity of an annihilation-powered source are connected \cite{gs69}.  If
$\Phi_{\gamma}$ is the photon flux from annihilations (photons\,cm$^{-2}$\,s$^{-1}$) and $\Phi_{\rm E}$ is the energy flux from
the same source (ergs\,cm$^{-2}$\,s$^{-1}$), then  $\Phi_{\gamma} \ga 10^{4}\,\Phi_{\rm E}$ \cite{gs69}.  Although annihilation
was proposed as a panacea for the energy budgets of QSOs and other high luminosity sources \cite{burbidgehoyle}, the detailed
emission mechanisms required enormous magnetic fields, compounding the problems of an already stretched energy budget.  Steigman
and Strittmatter \cite{gs71} explored whether observations of the annihilation neutrino flux could constrain models of annihilation-driven infrared emission in Seyfert galaxies.  For individual sources, it was estimated \cite{gs76} that the neutrino flux would be at least five orders of magnitude smaller than was observed at the time.  The difficulty of detecting the relatively low energy ($\la 500\,{\rm MeV}$) neutrinos, combined with improved models for the energy sources in QSOs, Seyferts, etc. have made annihilation neutrinos an unlikely probe.

Moving further away, outside our own galaxy, the strongest constraints, on the largest scales, come from observations of x-ray
emitting clusters of galaxies \cite{gs76,gs79,gs08}.  Most of the baryons in clusters of galaxies are in the hot intracluster
gas.  The same collisions between particles in the intracluster gas responsible for producing the observed x-ray emission would
result in annihilation gamma rays if some fraction of the gas consisted of antiparticles.  The virtue of using x-ray emitting
clusters of galaxies is that there is a direct proportionality between the x-ray emission from thermal bremsstrahlung and gamma
ray emission from annihilation.  This approach leads to bounds on the antimatter fraction (the fraction of antimatter mixed with
ordinary matter) on the largest scales in the Universe ($M \sim 10^{14} - 10^{15}\,M_{\odot}$, $R \sim {\rm few\,Mpc}$)
\cite{gs08}.  Using data from 55 x-ray emitting clusters of galaxies \cite{edge} in combination with the upper bounds to the
gamma ray fluxes \cite{egret}, it was found that the antimatter fraction from that sample is limited to $f <
10^{-6}$\,\cite{gs08}.  However, even stronger bounds exist for some individual clusters.  For the Perseus cluster, $f <
8\times10^{-9}$ and for the Virgo cluster, $f < 5\times10^{-9}$.  Perhaps the most interesting upper bound on antimatter on the
largest scales comes from colliding clusters.  Analysis of the Bullet Cluster gives $f < 3\times10^{-6}$ on the scale $M \sim
3\times10^{15}\,h^{-1}\,M_{\odot}$, where $h$ is the Hubble parameter in units of 100 km sec$^{-1}$ Mpc$^{-1}$ \cite{gs08}.

\subsection{The Problem of a Symmetric Universe}
\label{sec:annihilation}

Very shortly after the discovery of the CMB \cite{penzwil,dicke},
Ya.\,B.\,Zeldovich \cite{zeldovich}\,\footnote{It is interesting that
Zeldovich's article was written prior to the discovery of the CMB.  As a result,
in his review, Zeldovich considered both hot and cold universes.} and
H.\,Y.\,Chiu \cite{chiu}, independently, considered the fate of matter and
antimatter emerging from the early stages of the evolution of a hot Universe. 
The result, whose derivation is outlined here, is easily summarized.  At high
temperatures, above the quark - hadron transition, there are many
quark-antiquark pairs and, in a symmetric Universe, there are equal numbers of
quarks and antiquarks.  As the Universe expands and cools, the quarks (and
gluons) are confined into nucleons (neutrons and protons) which, because the
strong interaction is strong, are in thermal equilibrium with the cosmic plasma
(\eg photons, neutrinos, and the light leptons and bosons).  In this regime the
nucleon mass exceeds the temperature so that annihilation of nucleon-antinucleon
pairs proceeds on a timescale short compared to the expansion rate of the Universe.  But, since $m \gg T$, creation of new nucleon-antinucleon pairs from
collisions in the background plasma is strongly (exponentially) suppressed, so that up to spin-statistics factors of order unity, the ratio of nucleons
(and antinucleons) to photons is $n_{\rm N}/n_{\gamma} = n_{\bar{\rm N}}/n_{\gamma} = n_{eq}/n_{\gamma} \propto (m/T)^{3/2}e^{-(m/T)} \ll 1$.  Even though
the abundances of nucleons and antinucleons (\eg relative to photons) are very small, the strong interaction is strong, ensuring that $n_{\rm N} \approx
n_{eq}$ is maintained down to very low temperatures, $T \ll m_{\rm N}$.  However, eventually the abundance of the nucleon-antinucleon pairs becomes so
small that they no longer can find each other to annihilate (and the creation of new pairs is exponentially suppressed), and the abundance of nucleons (and
antinucleons) ``freezes out," at a ``relic" abundance $(n_{\rm N}/n_{\gamma})_{0}$.  The evolution of the nucleon-antinucleon abundances follows an
evolution equation, described next, that accounts for creation, annihilation, and the expansion of the Universe.  The solution, presented below, shows that
the relic abundance of the nucleon-antinucleon pairs in a symmetric Universe
is some nine orders of magnitude smaller than the nucleon abundance observed in our Universe, providing an important nail in the coffin of the symmetric Universe.

As first derived from an argument of detailed balance by Zeldovich \cite{zeldovich} and later rediscovered and supported by many textbook
derivations based on the Boltzmann equation, the evolution of the abundance of a particle (and its antiparticle) produced and annihilated in pairs,
is described by the standard evolution equation (SEE); see, \eg \cite{gs69,gs76,gs79,chiu,LW,ST,GG,SDB} and references therein.  For equal numbers of particles and antiparticles (no asymmetry, zero chemical potential), the SEE may be written as
\beq
{1 \over V}\bigg({dN \over dt}\bigg) = {dn \over dt} + 3Hn = \sigv(n_{eq}^{2} - n^{2})\,,
\label{eq:see}
\eeq
where $N = nV$ is the number of particles (and antiparticles) in a comoving volume $V$.  As the Universe expands and the cosmic scale factor, $a$,
increases, the comoving volume grows as $V \propto a^{3}$.  In Eq. (\ref{eq:see}), the number density of particles and antiparticles is $n$, the total annihilation cross section is \sigv, and $H = a^{-1}(da/dt)$ is the Hubble parameter.  The SEE is a form of the Ricatti equation, for which there are no known closed form
solutions except in special cases.  Although the SEE may be integrated numerically, here the approximate analytic approach first outlined by Zeldovich
\cite{zeldovich} and employed extensively in \cite{gs76,gs79,LW,ST,GG,SDB} and elsewhere, is followed.

For the approximate analytic solution to the SEE it is convenient to write $n = (1+\Delta)n_{eq}$ where, in the nonrelativistic (NR) regime ($T <
m$), $n_{eq} = (g\,T^{3}/(2\pi)^{3/2})x^{3/2}e^{-x}f(x)$, where $x \equiv m/T$ and $g = 2$ is the number of spin states of the proton (neutron) and
of the antiproton (antineutron).  Here $f(x)$ is an asymptotic series in $x$ for which  $f(x) \rightarrow 1$ as
$x \rightarrow \infty$.  For the range of $x$ of
interest in tracking the evolution of nucleon-antinucleon pairs, $f(x) \approx 1$ is a very good approximation.   Therefore, the evolution of the
equilibrium number density (as a function of $x$) in the NR regime is very well described by $n_{eq} \propto T^{3}x^{3/2}e^{-x} \propto
x^{-3/2}e^{-x}$.  Note that since the photon number density varies as $T^{3}$, $n_{eq}/n_{\gamma} \propto x^{3/2}e^{-x}$ in the NR regime.  Instead
of following the time evolution of the thermal relic abundance, it is more convenient to track its evolution as a function of $x$.  Neglecting small logarithmic corrections involving derivatives related to the entropy and photon densities, the derivatives with respect to time and $x$ (or $T$) are related by
\beq
dt \approx {1 \over H}\bigg({dx \over x}\bigg) \approx -{1 \over H}\bigg({dT \over T}\bigg)\,,
\eeq
where $H = H(T)$ is the Hubble parameter evaluated  at temperature $T$.
Now we define the quantity $g_{\rho}(T)$ (in analogy to $g_{s}$) by $g_{\rho}/2 \equiv
\rho/\rho_{\gamma}$, where $\rho$ is the {\it total} mass/energy density and
$\rho_{\gamma}$ is the energy density in photons alone.  
During those epochs in the evolution of the
Universe when the energy density is dominated by the contribution from relativistic particles (radiation dominated: RD),
$H \propto \rho_{\rm R}^{1/2} \propto g_{\rho}^{1/2}\rho_{\gamma} \propto g_{\rho}^{1/2}T^{2}$.
In terms of $\Delta$ and $x$, the SEE may be rewritten as
\beq
{d({\rm ln}(1+\Delta)N_{eq}) \over d({\rm ln}\,x)} = -\bigg({\Gamma_{eq} \over H}\bigg)\,y\,,
\label{eq:see2}
\eeq
where $\Gamma_{eq} \equiv \sigv n_{eq}$ and $y \equiv \Delta(2+\Delta)/(1+\Delta)$.

For $x \sim O(1)$ ($m \approx T$), $\Delta$ is very small and $n = n_{eq}$ is a very good approximation.  As the Universe expands and cools, $x$
increases and $\Delta$ grows exponentially (while $n_{eq}$ decreases exponentially), and the departure from equilibrium grows.
Define $x_*$ to be the value of $x$ for which
$\Delta(x_*) \equiv \Delta_* \sim O(1)$, so the true abundance, $n_{*}$, exceeds the equilibrium density, $n_{eq*}$, by  factor $1+\Delta_{*} > 1$.
(A more precise definition of $x_*$ is given below).
For $x \ga x_{*}$, $\Delta \ga \Delta_{*}$ and $n/n_{eq} > 1$ increases.  In this regime, where $n \ga n_{eq}$, the SEE simplifies,
\beq
dN/dt = \sigv(n_{eq}^{2} - n^{2})\,V \approx -\sigv n^{2}V = -\sigv N^{2}/V\,.
\eeq
This equation can be integrated directly from $t = t_{*}$ (when $T = T_{*}$ and $x = x_{*}$) to $t = t_{0}$ (when $T = T_{0} \ll T_{*}$ and $x \gg x_{*}$).  Replacing the evolution with time (or with $x$) by the evolution with temperature,
\beq
{dN \over N^{2}} \approx {\sigv \over VH}{dT \over T}\,,
\eeq
where the Hubble parameter varies as $H \approx H_{*}(T/T_{*})^{2}$ and the comoving volume increases with decreasing temperature as $V \approx V_{*}(T_{*}/T)^{3}$.  Integrating from $T = T_{*}$ to $T = T_{0} \ll T_{*}$ results in
\beq
N_{0}/N_{*} = [1 + (\Gamma/H)_{*}]^{-1}\,,
\eeq
where $\Gamma_{*} = n_{*}\sigv$.  For nucleon-antinucleon annihilation, $(\Gamma/H)_{*} \gg 1$, so that $N_{0}/N_{*} \approx (\Gamma/H)^{-1}_{*} \ll 1$.
When $T = T_{*}$ ($x = x_{*}$), the number of particles (neutrons or protons) in the comoving volume, $N_{*}$, may be compared to the number of photons in the same volume, $N_{\gamma*}$,
\beq
\bigg({N \over N_{\gamma}}\bigg)_{*} = \bigg({n \over n_{\gamma}}\bigg)_{*} = \bigg({H \over n_{\gamma}\sigv}\bigg)_{*}\bigg({\Gamma \over H}\bigg)_{*}\,.
\eeq
In terms of $x_{*}$,
\beq
\bigg({H \over n_{\gamma}\sigv}\bigg)_{*} = {6.5\times10^{-36}\,g_{\rho*}^{1/2}\,x_{*} \over m\sigv}\,,
\eeq
where $m$ is in GeV and \sigv~is in cm$^{3}$s$^{-1}$.
As the Universe expands and cools from $T = T_{*}$ to $T = T_{0}$, the surviving nucleon (and antinucleon) abundance(s) decrease to an asymptotic (``frozen out") value (ratio to photons) given by,
\beq
\bigg({N \over N_{\gamma}}\bigg)_{0} = \bigg({N \over N_{\gamma}}\bigg)_{*}\bigg({N_{0} \over N_{*}}\bigg)\bigg({N_{\gamma*} \over N_{\gamma0}}\bigg)\,,
\eeq
where, from entropy conservation, $N_{\gamma*}/N_{0} = g_{s0}/g_{s*}$.  Note that $(N/N_{\gamma})_{0}$ is the frozen out ratio of neutrons or protons to photons (long after annihilation has ceased\,\footnote{Annihilations never really cease.  They simply become so rare that they are unable to continue to reduce the relic abundance.}) and is identical to the ratio of antineutrons or antiprotons to photons.  Even though $(N/N_{\gamma})_{0} \neq 0$, the baryon asymmetry parameter in a symmetric Universe is $\etab = 0$.  Combining the above equations,
\beq
\bigg({N \over N_{\gamma}}\bigg)_{0} \approx {2.5\times10^{-35} \over m\sigv}\bigg({g_{\rho*}^{1/2} \over g_{s*}}\bigg)x_{*}\,,
\eeq
where $g_{s0} = 43/11$, corresponding to $N_{\nu} = 3$, has been adopted.  For neutrons or protons (in the approximation here they are assumed to have the same mass, $m \approx 0.94\,{\rm GeV}$), the total (s-wave) annihilation cross section\,\footnote{Even though $T_{*} \ll m$, the nucleons are moving sufficiently rapidly that Coulomb (Sommerfeld) enhancement of the proton-antiproton annihilation cross section, relative to the neutron-antineutron annihilation cross section, is unimportant.} is $\sigv \approx 1.5\times10^{-15}\,{\rm cm^{3}\,s^{-1}}$, so that
\beq
(N/N_{\gamma})_{0} \approx 1.8\times10^{-20}(g_{\rho*}^{1/2}/g_{s*})x_{*}\,.
\label{eq:n0}
\eeq
To find $x_{*}$ and $T_{*} = m/x_{*}$, in order to evaluate $g_{\rho*} = g_{\rho}(T_{*})$ and $g_{s*} = g_{s}(T_{*})$,\footnote{For $g_{\rho}(T)$
and $g_{s}(T)$, the results of Laine and Schroeder \cite{ls} are used here.}, we impose the condition defining $x_{*}$, that is, when $x = x_{*}$,
$\Delta(x) = \Delta(x_{*}) \equiv \Delta_{*}$.  Although $\Delta_{*} \sim O(1)$, a specific choice needs to be made for $\Delta_{*}$ in order to find the corresponding value of $x_{*}$ (and it needs to be checked and confirmed that the final result is insensitive to this specific choice).  Here, $\Delta_{*} = 0.618$ (related to the ``Golden Mean") is adopted, so that $y_{*} = \Delta_{*}(2+\Delta_{*})/(1+\Delta_{*}) = 1$.

It may be verified that $d({\rm ln}(1+\Delta))/d({\rm ln}\,x) \ll d({\rm
ln}\,N_{eq})/d({\rm ln}\,x)$, so that Eq.\,(\ref{eq:see2}) reduces to
\beq
-(\Gamma_{eq}/H)\,y \approx d({\rm ln}\,N_{eq})/d({\rm ln}\,x)\,,
\eeq
where $N_{eq} = n_{eq}V \propto VT^{3}x^{3/2}e^{-x}$.  Generally, $VT^{3}
\propto (aT)^{3} \approx {\rm constant}$, so that the logarithmic derivative of
$VT^{3}$, depending on $d({\rm ln}\,g_{s})/d({\rm ln}\,dT)$, may be neglected,
further simplifying Eq.\,(\ref{eq:see2}) to an algebraic equation, 
\beq
d({\rm ln}\,N_{eq})/d({\rm ln}\,x) \approx -(x - 3/2) \approx -(\Gamma_{eq}/H)\,y\,.
\eeq
For $x = x_{*}$, $\Delta(x_{*}) = \Delta_{*} = 0.618$ and $y = y_{*} = 1$.  As a result,
\beq
x_{*} - 3/2 = (\Gamma_{eq}/H)_{*} = n_{eq*}\sigv/H_{*} = A_{*}\,g_{\rho*}^{-1/2}x_{*}^{1/2}e^{-x_{*}}\,,
\label{eq:x*}
\eeq
where $A_{*} \equiv 4\times10^{34}g\,m\sigv$; $g$ is the number of neutron or
proton spin states and, as before, the mass $m$ is in GeV and \sigv~is in
cm$^{3}$/s.  For $g = 2$, $m = 0.94$, and $\sigv = 1.5\times10^{-15}$, $A_{*} = 1.1\times10^{20}$.  The transcendental equation for $x_{*}$, Eq.
(\ref{eq:x*}), may be solved iteratively.  The solution is $x_{*} \approx 43.1$, corresponding to $T_{*} = m/x_{*} \approx  21.8\,{\rm MeV}$, for
which $g_{\rho*} \approx 11.5$ and $g_{s*} \approx 11.4$ \cite{ls}.  Substituting these values into Eq. (\ref{eq:n0}) results in the frozen-out ratios of the surviving numbers of neutrons, protons, antineutrons, and antiprotons to photons,
\beq
\bigg({n_{\rm n} \over n_{\gamma}}\bigg)_{0} = \bigg({n_{\rm p} \over n_{\gamma}}\bigg)_{0} = \bigg({n_{\bar{\rm n}} \over n_{\gamma}}\bigg)_{0} = \bigg({n_{\bar{\rm p}} \over n_{\gamma}}\bigg)_{0} \approx 2.3\times10^{-19}\,.
\eeq
The corresponding nucleon (neutron plus proton) and antinucleon to photon ratios are $(n_{\rm N}/n_{\gamma})_{0} = (n_{\bar{\rm N}}/n_{\gamma})_{0} \approx 4.6\times10^{-19}$.  

Of course, for this symmetric Universe, $\etab \equiv (n_{\rm N}/n_{\gamma})_{0} - (n_{\bar{\rm N}}/n_{\gamma})_{0} = 0$.  The present
mass density of matter (nucleon plus antinucleon) is $\rho_{\rm B} = m(n_{\rm N} + n_{\bar{\rm N}}) \approx 3.5\times10^{-16}\,{\rm
GeV\,cm^{-3}}$, or $\omhhb \approx 3.3\times10^{-11}$.  In contrast, for our observed {\it asymmetric} Universe, where annihilation of any relic antinucleons is very efficient, $(n_{\rm N}/n_{\gamma})_{0} \approx 6.1\times10^{-10} \gg (n_{\bar{\rm N}}/n_{\gamma})_{0} \approx 0$ and $\omhhb \approx 0.022$.  In a symmetric Universe the abundance of nucleons surviving annihilation in the early Universe is smaller than the abundance of nucleons in our asymmetric Universe by some nine orders of magnitude.

Notice that when $T = T_{*}$, the ratio of the annihilation rate to the
expansion rate is very large, $(\Gamma/H)_{*} \approx (1+\Delta_{*})(x_{*} - 3/2) \approx 67 \gg 1$.
Neither annihilations nor the relic abundances freeze out when $T = T_{*}$.  For $T < T_{*}$, annihilations continue to reduce the abundances of nucleons and antinucleons and the ratio of the annihilation rate to the expansion rate, $\Gamma/H$, continues to decrease.  Eventually, for $T \equiv T_{f} \approx T_{*}/2$, $(\Gamma/H)_{f} = 1$, and the relic abundances freeze out (although, depending on $T_{f}$, the number of photons in the comoving volume may continue to increase until $T \la m_{e}$, further reducing the relic baryon to photon ratio).  For $T < T_{f}$, $n_{\rm N} = n_{\rm \bar{N}} = n_{{\rm N}f}\,(T/T_{f})^{3}$.  For temperatures even slightly below $T_{f}$, $(\Gamma/H) \approx H_{f}/H = (T_{f}/T)^{2} < 1$.  Thereafter, the annihilation rate scales as $n\sigv \propto T^{3}$ (for s-wave annihilation), while the expansion rate of the Universe scales as $H \propto T^{2}$ (during radiation dominated epochs in the evolution), so that after freeze out ($T \ll T_{f}$), $\Gamma/H \approx T/T_{f} \ll 1$.  During matter dominated epochs in the evolution of the Universe, $H \propto T^{3/2}$, so that $\Gamma/H \propto T^{3/2}$, and it is still the case that $\Gamma/H \ll 1$.

By the same argument, nuclear reactions in this Universe are extremely suppressed by the very low nucleon density.  There can be no primordial
nucleosynthesis in a symmetric Universe.  After freeze out, as the Universe expands and cools, neutrons decay and the Universe is left with protons
(and antiprotons) and electrons (and positrons).  Note that as the protons and electrons (and antiprotons and positrons) cool and become
nonrelativistic, the long-range Coulomb interaction enhances, through Sommerfeld enhancement \cite{sommerfeld}, the annihilation cross section, $\sigv \rightarrow 2\pi(\alpha\,c/v)\sigv \propto T^{-1/2}$.  Even so, the ratio of the annihilation rate to the expansion rate still decreases (as $T^{1/2}$ during RD epochs and as $T$ during MD epochs).  Recombination cannot occur in such a low baryon density Universe.  In the absence of non-baryonic dark matter, it is unlikely that any collapsed structures (\eg stars or galaxies) could form in such a low density, ionized Universe.  The history (and future) of a symmetric Universe is very bleak.  The story barely changes if a symmetric Universe contains non-baryonic dark matter.  If, for example, the presence of DM in a symmetric Universe allows collapsed DM structures to form, the relic matter and antimatter would fall into the DM potential wells, increasing their number densities, leading to renewed annihilation, further reducing their already very small abundances.  A matter-antimatter symmetric Universe simply bears no resemblance to our Universe.\\

Even in an asymmetric Universe, during the very early evolution of the Universe when the temperature is very high, the {\it equilibrium} abundance of nucleons and antinucleons may be much larger than the relic abundance of nucleons in our Universe, $\etab = (n_{\rm N}/n_{\gamma})_{0} \approx 6\times10^{-10}$.  These pairs will annihilate until, at some temperature, $T$, $(n_{\rm N}/n_{\gamma})(g_{s}(T)/g_{s0}) \approx 6\times10^{-10}$.  For lower temperatures, the antinucleons continue to be annihilated but the nucleons, due to the asymmetry, are frozen out.  For nucleons (protons plus neutrons), $g = 4$, and their equilibrium abundance relative to photons is $n_{\rm N}/n_{\gamma} = 0.26\,g\,x^{3/2}e^{-x} \approx x^{3/2}e^{-x}$, where $x = m_{\rm B}/T$ and, prior to BBN, the average mass per baryon is $m_{\rm B} \approx 939$ MeV \cite{gs06}, so that $x \approx 939/T$, with $T$ in MeV.  Here, we have assumed that $f(x) \approx 1$.  To find $T$, we need to solve $(939/T)^{3/2}{\rm exp}(-939/T) \approx 1.5\times10^{-10}g_{s}(T)$.  Using \cite{ls} for $g_{s}(T)$, the solution is $T \approx 38\,{\rm MeV}$ ($x \approx 25$).  To avoid the annihilation catastrophe in a symmetric Universe, the baryon asymmetry must have been created when $T > 38\,{\rm MeV}$ (or, when $T \gg 38\,{\rm MeV}$).  Recall that $T_{*} \approx 22\,{\rm MeV}$, so $T > T_{*}$, as expected.  In the extensions of the standard models of particle physics and cosmology that allow for a baryon asymmetry at low temperatures, the energy/temperature/mass scales are orders of magnitude larger than this conservative estimate.

\section{Particle Physics Models for Generating the Universal Matter-Antimatter Asymmetry}
\label{sec:sakharovetal}

It is clear from the preceding two sections that a Universe containing equal abundances of baryons and antibaryons
is {\it not} the Universe we actually observe.  At some point in its evolution, the Universe most have developed an asymmetry
between matter and antimatter. How did this asymmetry come about?

One possibility is that the Universe actually began in an asymmetric state, with more baryons and antibaryons.  This is,
however, a very unsatisfying explanation.  Furthermore, if the Universe underwent a period of inflation (i.e., very rapid
expansion followed by reheating), then any preexisting net baryon number would have been erased.  A more natural
explanation is that the Universe began in an initally symmetric state, with equal numbers of baryons and antibaryons,
and that it evolved later to produce a net baryon asymmetry.

As we noted in the introduction, Sakharov introduced three conditions necessary to produce a net baryon asymmetry in a
Universe that began
with zero net baryon number.  These Sakharov conditions form the basis of nearly all modern theories of baryogenesis,
so we will review them in more detail here.  These conditions are:

\vskip 0.5 cm

\noindent 1.  Baryon number violation.  This is the most obvious component needed for baryogenesis.  If the universe began with zero net baryon number, and
baryon number were conserved, then it would still have zero net baryon number today.

\vskip 0.5 cm

\noindent 2.  C and CP violation.  The operator C changes particles into antiparticles and vice versa, while CP also flips all three
coordinate axes.  A universe that is baryon-antibaryon symmetric is unchanged when C or CP is applied, while the same is not true for a universe
with a net baryon excess.  Hence, the production of a baryon asymmetry requires C and CP violation.

\vskip 0.5 cm

\noindent 3.  A departure from thermodynamic equilibrium.  If baryon and C/CP were violated while thermal equilibrium conditions prevailed, then the chemical
potentials for baryons would be driven to zero, and the only possible difference between particle and antiparticle abundances would arise if there
were a mass difference between them.  But CPT invariance implies that the masses of particles and antiparticles are the same.  Hence, Sakharov conditions 1
and 2 allow for a net baryon number to be created only when the particles of interest are out of thermal equilibrium.

\vskip 0.5 cm

While we know the general conditions necessary to generate a baryon
asymmetry from an initially symmetric state, we are far from having a single
accepted theory of baryogenesis.  Here we will outline some of the ideas
that have been proposed over the years.  For some of the earliest work in this field, see
Refs. \cite{susskind,yoshimura,ttwz,weinberg,ellis}.  For reviews of this topic, see
Refs. \cite{Riotto,Dine}.

Perhaps the simplest class of models (and one of the earliest to be investigated) involves the decay of massive particles.
Consider a particle-antiparticle pair, $X$ and $\bar X$, that has dropped out of thermal equilibrium in the early Universe, in the sense
defined in \S\ref{sec:annihilation}.  Suppose the $X$ can decay into two different channels, with baryon numbers $B_1$ and $B_2$, respectively,
while $\bar X$ decays into the corresponding ``anti"-channels, with baryon numbers $-B_1$ and $-B_2$, respectively.  Invariance under CPT
guarantees that the {\it total} decay rate for an antiparticle must be equal to the decay rate for the corresponding particle.  However, it
says nothing about individual branching ratios.  So it is possible, for instance, for the branching ratio of $X$ into the channel with baryon
number $B_1$ (which we will take to be $r$) to be different from the branching ratio of $\bar X$ into the channel with baryon number $-B_1$,
which we will call $\bar r$.  The possibility of such a difference
is the key idea underlying this mechanism for baryogenesis. Note that $r \ne \bar r$ is only possible
if C and CP are violated.

With the branching ratios and baryon numbers defined above, the net baryon number
produced from each pair of $X$ and $\bar X$ decays is:
\begin{eqnarray}
B &=& B_1 r + B_2 (1-r) - B_1 \bar r - B_2 (1 - \bar r),\nonumber \\
\label{baryon}
&=& (B_1 - B_2)(r - \bar r).
\end{eqnarray}
Eq. (\ref{baryon}) illustrates the necessity of the three Sakharov conditions.  If C and CP were not violated, we would have $r = \bar r$, and the
right-hand side of Eq. (\ref{baryon}) would be zero.  Similarly, the possibility that $X$ can decay into two different channels with
different baryon numbers is only possible if $B$ is not conserved; otherwise we would have $B_1 = B_2$ and again the right-hand side of
Eq. (\ref{baryon}) would be zero.  Finally, we assumed out-of-equilibrium conditions in setting up this scenario, i.e., when they decay, $X$ and $\bar X$
are not in equilibrium with the thermal background, either through annihilations with each other or through inverse decays.
If this were not the case, the particles produced in the $X$ and $\bar X$ decays would simply assume thermal equilibrium abundances,
which would yield equal baryon and antibaryon densities.

The scenario we have sketched out here is a toy model; for more detailed models see, e.g., Ref. \cite{KT}.  Models of this sort were first
advanced in connection with physics at the GUT (grand-unified) scale, $T \sim 10^{15} - 10^{16}$ GeV.  However, these ideas run into trouble
if inflation is assumed to occur in the early universe.  The reason is that, as we have noted, inflation wipes out any preexisting baryon asymmetry, so that
baryogenesis must occur after inflation, and currently-favored models of inflation do not reheat the universe to a temperature
as high as the GUT scale.

Another possibility for baryogenesis is the Affleck-Dine mechanism \cite{ad}.  This model is motivated by supersymmetry, in which all
of the particles of the Standard Model have corresponding superpartners with opposite spin statistics (fermions are paired with bosonic
superparticles and bosons with fermionic superpartners).  The Affleck-Dine mechanism invokes a scalar field that can carry a net baryon number.  The field is
initially frozen at early times, but begins oscillating when the Hubble parameter drops below its mass.  During these oscillations, the scalar
field acquires a net baryon number, which is transferred at later times into standard model particles.

Electroweak baryogenesis \cite{Kuzmin} is based on the idea that the universe underwent an electroweak phase transition at a temperature $T \sim 100$ GeV,
when the Higgs field dropped into its vaccuum state, giving masses to the quarks, leptons, and gauge bosons.
If the electroweak phase transition is first order, it can temporarily drive the universe out of thermal equilibrium as bubbles of the low-temperature vacuum nucleate,
expand, and collide, ultimately occupying all of space.  The production of baryons
occurs in this out-of-equilibrium state near the walls of these expanding bubbles.  Electroweak baryogenesis does require physics beyond the standard model, as
the measured Higgs boson mass implies that the phase transition would not be
first order in the standard model.  This new physics would couple to
the Higgs boson, altering its production and decay.  Thus, the viability of these models can be
tested in the laboratory.

Another proposal goes under the heading of leptogenesis \cite{fukugita}.  These models
are based on a result by 't Hooft \cite{thooft}, who showed that even
in the Standard Model, baryon number is violated by
nonperturbative electroweak processes.  These processes conserve B$-$L, but not B and L separately.
Furthermore, while the rates for such processes are very low at low temperatures, they can be much higher
in the early universe.  Leptogenesis then, is the production of
a net lepton asymmetry
in the early universe, e.g., through massive particle decay as discussed above.  Then nonperturbative electroweak effects
transfer some of the net
lepton number into a net baryon number. 

This is by no means an exhaustive list of models for baryogenesis, which remains very much an open and active field of research.  At this
point we are confident of the ingredients required in any successful model (the Sakharov conditions), and we have a very accurate
measure of the desired outcome (the observed baryon asymmetry), but the determination of the correct model for baryogenesis
remains an ongoing effort.

\section{The Baryon Asymmetry Parameter and Primordial Nucleosynthesis}

\label{sec:bbn}

\begin{figure}
\begin{center}
\includegraphics[width=0.85\columnwidth]{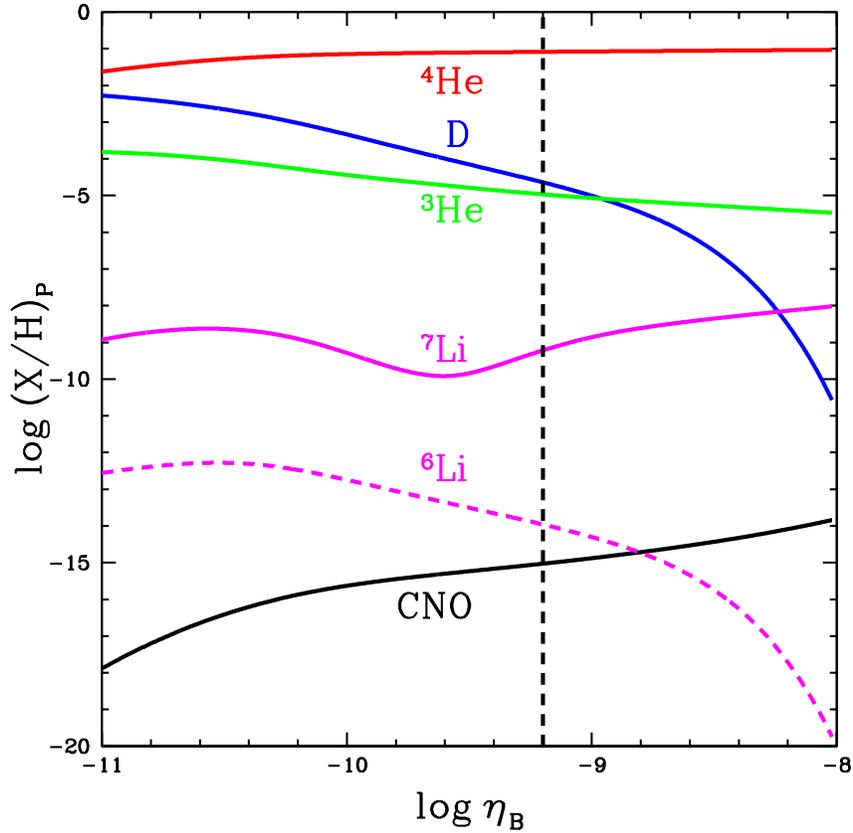}
\caption{The primordial abundances predicted by SBBN \cite{KMN}  for a large range of the present value of the baryon to photon ratio $\etab = (n_{\rm B}/n_{\gamma})_{0}$.  For all abundances (including \4he) the ratio to hydrogen by number is shown.  The dashed vertical line indicates the current SM value of $\etab \approx 6\times10^{-10}$.}
\label{fig:sbbn}
\end{center}
\end{figure}

In the standard model of particle physics and cosmology, the baryon asymmetry parameter plays a key role in BBN, regulating the rates of the nuclear reactions synthesizing (and destroying) the nuclides heavier than hydrogen.  BBN in the standard model (SBBN) and in extensions of the SM when various nuclear physics and other parameters are allowed to vary is described in Uzan's contribution to this volume.  Here we are mainly concerned with the BBN predicted primordial (prestellar) abundances of the light nuclides, along with the CNO abundances.

\subsection{Standard BBN}

The SBBN predicted abundances \cite{KMN}, the ratios by number compared to hydrogen, are shown as a function of \etab~in Figure \ref{fig:sbbn} for a factor of 1000 range in \etab, for the SM case of $N_{\nu} = 3$.
Agreement between the predicted and the observationally-inferred deuterium abundance
and the Planck observations of the CMB power spectrum imply a value of $\etab \sim 6\times10^{-10}$.
This value is shown by the dashed vertical line in Fig.\,\ref{fig:sbbn}.
This value of $\eta_B$ also provides good agreement with the primordial $^4$He abundance derived from observations.  However, it predicts
a primordial $^7$Li abundance roughly three times larger than the observationally-inferred abundance; this primordial
lithium problem remains unresolved at present (see Ref. \cite{Fields} for a recent review).

As seen in Fig.\,\ref{fig:sbbn}, over this large range in the baryon asymmetry parameter, the abundance trends are quite simple: as \etab~increases, the \4he~abundance increases monotonically, but very slowly ($\sim$ logarithmically); the abundances of D, \3he, and \6li are all monotonically decreasing, while the abundances of the CNO nuclides increase.  In contrast, the evolution of the abundance of \7li~is non-monotonic.  Starting from very small values of \etab, as \etab~increases, the \7li~abundance first increases (until $\etab \sim 3\times10^{-11}$), then decreases (until $\etab \sim 3\times10^{-10}$), and finally increases again (eventually, for even larger values of \etab, the \7li~abundance will decrease, being replaced by CNO and heavier nuclides).  As \etab~increases from $10^{-11}$ to $10^{-8}$, \4he/H~increases by a factor of $\sim 4$, from \4he/H~$\sim 0.024$ (\Yp~$\sim 0.09$) to \4he/H~$\sim 0.093$ (\Yp~$\sim 0.27$) and the deuterium abundance decreases dramatically, from D/H $\sim 5\times10^{-3}$ to D/H $\sim 3\times10^{-11}$.  Over the same range in \etab~the \3he~abundance decreases more slowly, from $\sim 10^{-4}$ to $\sim 3\times10^{-6}$ and the \7li abundance ranges from $\ga 10^{-10}$ to $\la 10^{-8}$, while the abundance of the CNO nuclides increases from $\sim 10^{-18}$ to $\sim 10^{-14}$.

For the value of the baryon asymmetry parameter inferred for the observed Universe, $\etab \sim 6\times10^{-10}$, \4he/H~$\sim 0.082$ (\Yp~$\sim 0.25$), D/H $\sim 2.5\times10^{-5}$, \3he/H $\sim 1.1\times10^{-5}$, \7li/H $\sim 5.4\times10^{-10}$, and the abundances of all the other primordial nuclides are $\la 10^{-14}$.  For a very wide range in the baryon asymmetry parameter, the gas that will become the first stars in the Universe consists mainly of hydrogen and helium (\4he), with only trace amounts of any other, heavier nuclides.  Note, however, that for $\etab \gg 10^{-8}$, the primordial abundances of the CNO and heavier nuclides may become non-negligible (see \S\,\ref{sec:wfh} below).  In the absence of significant CNO (or D) abundances, it is the hydrogen and helium content of the primordial gas that will most influence the formation, structure, and evolution of the first stars.\\

\subsection{BBN for a Larger Range of Baryon Asymmetries}
\label{sec:wfh}

In the seminal BBN paper of Wagoner, Fowler, and Hoyle (WFH)\,\cite{wfh}, and in several follow up papers by Wagoner \cite{wagoner73}, Schramm and Wagoner \cite{schrammwagoner}, and Schramm \cite{schramm98}, a much larger range in the baryon asymmetry parameter was explored than is shown here in Figure\,\ref{fig:sbbn}.  In the WFH paper a range of some eight and a half orders of magnitude was considered, $-12 \la {\rm log}\,\etab \la -3.5$, while in the other cited papers the range is five orders of magnitude, $-11 \la {\rm log}\,\etab \la -6$.  Although the quantitative BBN yields in those papers, based on what are now outdated nuclear and weak interaction rates (especially the much revised neutron lifetime), should be taken with a large grain of salt, the trends of the yields with \etab~revealed in those papers are likely robust.  

For example, over the entire range explored, the helium mass fraction increases (and the hydrogen mass fraction decreases) monotonically
with \etab.  Over the same range in \etab, the D and \3he~mass fractions decrease monotonically, with the deuterium abundance falling
much more rapidly than the \3he~abundance.  The evolution of the \7li~mass fraction, $X_{7}$, is more interesting.  At the lowest baryon
asymmetries, $X_{7}$ increases from being negligible at ${\rm log}\,\etab \sim -12$ to a local maximum, a hint of which may be seen in
Fig.\,\ref{fig:sbbn}, when ${\rm log}\,\etab \sim -10.5$.  Then, as \etab~continues to increase, $X_{7}$ decreases to a local minimum at
${\rm log}\,\etab \sim -9.5$, as may be seen in Fig.\,\ref{fig:sbbn}.  For $\eta_B \ga 3 \times 10^{-10}$, $X_{7}$ increases to another local maximum when ${\rm log}\,\etab \sim -6$, after which $X_{7}$ decreases monotonically for all larger values of \etab.  For ${\rm log}\,\etab \la -8$, the abundances of the CNO and heavier nuclides are negligible.  As \etab~continues to increase, so too, do the CNO abundances, surpassing the \3he and \7li abundances for ${\rm log}\,\etab \ga -6$.  However, almost as soon as the CNO nuclides become large enough to be of possible interest, they decrease as \etab~continues to increase, being replaced by even heavier nuclides.  The trend seen at the very highest values of the baryon asymmetry parameter in the WFH paper suggests that at sufficiently high values of \etab~the iron peak elements might be produced during primordial nucleosynthesis.  It is interesting to speculate if even larger baryon to photon ratios might lead to the r-process elements.

As discussed in \S\ref{sec:groundrules}, in determining if the baryon asymmetry parameter is fined tuned, we are asking if stars, planets, and life
could exist in alternate universes
with different values of $\eta_B$.  In this case we need to check if the primordial abundances in alternate universes allow for the cooling and collapse of primordial gas clouds to form the first stars, and if in the course of evolution of those stars, the elements required for life can be synthesized.\\

\section{Relation Between the Baryon Asymmetry Parameter and the Observable Cosmological Parameters}
\label{sec:cosmo}

In our present-day Universe, the parameter $\eta_B$ is not a directly observable quantity.  Instead, we measure quantities
such as the baryon density or the CMB temperature, from which $\eta_B$ can be inferred.  In this section we examine
the relation between $\eta_B$ and the observable cosmological quantities.

In a matter-antimatter asymmetric Universe such as ours, the baryon asymmetry parameter is related to the contribution of baryons (normal matter) to the total mass density.  As a result, the magnitude of the baryon asymmetry plays a role in the evolution of the Universe and in the growth and evolution of structure in it.  For the discussion here is it assumed that the Universe is, on average, homogeneous and is expanding isotropically, so that its evolution is described by the  ``Friedman equation",
\beq
\bigg({H \over H_{0}}\bigg)^{2} = \omr\bigg({a_{0} \over a}\bigg)^{4} +\,\omb\bigg({a_{0} \over a}\bigg)^{3} +\,\omdm\bigg({a_{0} \over a}\bigg)^{3} +\,\omk\bigg({a_{0} \over a}\bigg)^{2} + \oml\,.
\label{eq:friedman}
\eeq
In Eq. (\ref{eq:friedman}) the subscript $0$ indicates the present ($t = t_{0}$) value
of the parameters and $H = a^{-1}(da/dt)$ is the Hubble parameter, quantifying the expansion rate of the Universe,
where $a = a(t)$ is the cosmic scale factor.  The subscripts, R, B, DM, k, and $\Lambda$ stand, respectively,
for the contributions to the total mass/energy density from ``radiation" (i.e., massless particles or particles whose total energy (rest mass plus
kinetic)  far exceeds the rest mass energy), baryons (``normal" or ``ordinary" matter), dark matter (nonbaryonic
matter),\footnote{For agreement
with observations of structure formation and its growth as the Universe evolves, it is assumed that the DM is ``cold", in the sense that for those
epochs when deviations from homogeneity occur, the DM particles are moving
slowly ($v \ll c$).} curvature, and a cosmological constant.
For simplicity, we assume here that the observed accelerated expansion
of the universe is driven by a cosmological constant rather
than a time-varying dark energy component.
Since the mass densities of baryonic and dark matter evolve the same way (\eg $\rho \propto a^{-3}$), it is convenient to introduce a
parameter describing the ``matter density", the total mass density in nonrelativistic particles, $\omm \equiv \omb +\,\omdm$.  At the present epoch
($t = t_{0}$) a ``critical density" of the Universe may be identified, $\rho_{crit\,0} \equiv 3H_{0}^{2}/8\pi G = 1.05\times10^{-5}h^{2}\,{\rm
GeV\,cm^{-3}}$, where $H_{0} \equiv 100\,h\,{\rm km\,s^{-1}\,Mpc^{-1}}$ and $G$ is Newton's gravitational constant.  Here, and elsewhere, we will
often set $c = 1$ and express masses in energy units.  The parameters $\Omega_{i}$ that appear in Eq.\,(\ref{eq:friedman}) are the ratios of the various contributions to the present energy densities, normalized to the present critical density: $\Omega_{i} \equiv (\rho_{i}/\rho_{crit})_{0}$.

Consider the relation between the baryon asymmetry parameter (\etab), the baryon mass density parameter (\omb), and the Hubble constant ($H_{0}$).  The present mass/energy density in ordinary (baryonic) matter is $\rho_{\rm B0} = m_{\rm B}n_{\rm B0} = \omb\,\rho_{crit\,0} \approx 1.05\times10^{-5}\,\omhhb\,{\rm GeV\,cm^{-3}}$.  For an average mass per baryon of $m_{\rm B} \approx 0.938\,{\rm MeV}$\,\cite{gs06}\,\footnote{In the post-BBN Universe, when the baryons are mainly protons and alpha particles (hydrogen and helium), the average mass per baryon depends on the helium abundance (mass fraction, \Yp).  For \Yp~$\approx 0.25$, $m_{\rm B} \approx 938.112 + 6.683($\Yp$~- 0.250)\,{\rm MeV}$\,\cite{gs06}.}, the present baryon number density is $n_{\rm B0} \approx 1.12\times10^{-5}\,\omhhb\,{\rm cm^{-3}}$.  If the present temperature of the CMB photons
is $T_{0}$ (in degrees Kelvin), then the present photon number density is $n_{\gamma0} \approx 20.3\,T_{0}^{3}\,{\rm cm^{-3}}$, and the
present baryon to photon ratio is $\etab \approx 5.54\times10^{-7}\,(\omhhb/T_{0}^{3})$, so that $\eta_B \approx 5.54\times10^{-7}\,(\omhhb/T_{0}^{3})$ and $n_{\rm B}/s \approx 7.87\times10^{-8}\,(\omhhb/T_{0}^{3})$ (for three flavors of SM neutrinos).  Note that the connection between the baryon asymmetry parameter (\etab~or $n_{\rm B}/s$) and the present mass density in ordinary matter ($\propto \omhhb$) depends on the present ($t = t_{0}$) value of the photon temperature ($T_{0}$).  If the baryon asymmetry parameter were to change by some factor, the combination $\omhhb/T_{0}^{3}$ would change by the same factor, resulting in changes to the other universal observables (\eg \omb,\,$h$, $T_{0}$), separately or in combination.  The baryon asymmetry parameter is degenerate with these other cosmological parameters.  In particular, changes in \omb~alone would change the expansion history of the Universe, as may be seen from the Friedman equation.  The interconnections (degeneracies) among the cosmological observables complicates any discussion of the effect on the history and evolution of the Universe resulting from changes to any one of them (\eg the baryon asymmetry parameter).

If the Friedman equation, Eq. (\ref{eq:friedman}), is evaluated at present ($t = t_{0}$), when $H = H_{0}$, there is one condition on the five parameters,
\beq
1 = \omr +\,\omb + \,\omdm + \,\omk +\,\oml = \omr +\,\omm +\,\omk + \,\oml\,,
\eeq
leaving four free parameters.  For our observed Universe $\omk \ll 1$ and $\omr \ll 1$, so that $\omb +\,\omdm +\,\oml \approx 1$.  There are
still three parameters and only one constraint, leaving two free parameters.  By writing $\omm = \omb +\,\omdm$, it might appear that there are
only two parameters and one constraint, $\omm +\,\oml \approx 1$.  However, the ratio $\omb/\omdm$ remains a free parameter, so there are still
three parameters with one constraint among them.

In the next section we will consider how the evolution of the universe changes
when $\eta_B$ differs from its observed value.  While our intention is
to keep all of the other cosmological parameters constant, there remains
an ambiguity in the way we treat them.  Note that $\eta_B$ is a dimensionless
ratio of two quantities, the baryon and photon number densities.  When
we alter this quantity, we can consider two different possibilities:  (1)
changing $n_B$ relative to the other cosmological parameters, while leaving $n_\gamma$ unchanged relative
to these parameters, or
(2) keeping
$n_B$ fixed while changing $n_\gamma$ relative to the other cosmological parameters.
While each of these possibilities produces
a change in $\eta_B$, they differ in their treatment of the way that
$n_B$ and $n_\gamma$ change relative to the other cosmological quantities
of interest.  (Of course, these are only the two simplest possibilities; one could consider allowing the ratios of
{\it both} $n_B$ and $n_\gamma$ relative to the other cosmological parameters to change, but by different amounts, thus
changing $\eta_B$ as well).

Which of the two approaches spelled out in the previous paragraph is the correct one?  Absent a particular model for a
different universe with a different value of $\eta_B$, it is impossible to say.
However, the first possibility seems the more natural one. If we assume
that baryogenesis is independent of the processes that led to dark matter
or dark energy, then tweaking the model for baryogenesis will alter $n_B$
by the same factor relative to all of the other cosmological parameters
of interest.  This is the case we will consider in detail.

Let $F$ be the ratio of the value of $\eta_B$ in some hypothetical Universe
relative to its value in our Universe; our goal will be to understand
what constraints, if any, can be placed on $F$.  We will use a tilde to denote
physical quantities in a hypothetical universe in which $\eta_B$
has changed, and quantities without a tilde will denote the corresponding
values of these quantities in our universe, so that
\begin{equation}
\widetilde \eta_B = F \eta_B.
\end{equation}
In case (1) discussed above, the ratios $\rho_B/\rho_{DM}$,
$\rho_B/\rho_{\Lambda}$, and $n_B/n_\nu$ change in proportion to the change in $\eta_B$,
while $n_\gamma/\rho_{DM}$, $n_\gamma/\rho_{\Lambda}$, and $n_\nu/n_\gamma$ remain the same.
Thus, we have
\begin{eqnarray}
\label{change1}
\widetilde \rho_B/\widetilde \rho_{DM} &=& F \rho_B/ \rho_{DM},\\
\label{change2}
\widetilde \rho_B/\widetilde \rho_{\Lambda} &=& F  \rho_B/ \rho_{\Lambda},\\
\label{change3}
\widetilde n_B/\widetilde n_\nu &=& F n_B/n_\nu.
\end{eqnarray}

Of course, there are other possibilities that we will not explore here.  In an alternate universe with a late
production of entropy, $n_B$
would remain unchanged, while the
ratio of $n_\gamma$ to all of the other cosmological parameters would be altered.
Alternately, if baryogenesis were linked to the process that produced dark matter (as it is
in some models), one might consider the possibility of changing $\eta_B$ while leaving
$\rho_B/\rho_{DM}$ fixed.  Nonetheless, we feel that the model spelled
out in Eqs. (\ref{change1}) - (\ref{change3}) is the most natural way in which to
modify $\eta_B$, and this is the case we will now attempt to constrain.

\section{Alternate Universes with Different Baryon Asymmetry Parameters}
\label{sec:alternate}

Changing $\eta_B$ alters the evolution of the universe in two ways:  it changes
BBN, and it alters the processes that give rise to structure formation and
ultimately yield stars and planets.  We will consider both effects in turn.

First, consider our Universe at present.
Our Universe is very well described by a $\Lambda$CDM cosmological model with
$\omk \approx 0$, $\omr \ll 1$, and $\omb < \omdm < \oml$ ($\omb\,+\,\omdm\,+\,\oml \approx 1$).  For our observed Universe, a good approximation to the 2015 Planck CMB observations \cite{planck} is $\oml \approx 0.7$, $\omm \approx 0.3$, $\omb \approx 0.05$, $\omdm \approx 0.25$.  For a $\Lambda$CDM cosmology with $\oml \approx 0.7$, $H_{0}t_{0} \approx 0.96$, so that for $H_{0} \approx 68\,{\rm km\,s^{-1}\,Mpc^{-1}}$, $t_{0} \approx 13.8\,{\rm Gyr}$.  For the present CMB temperature, the Fixsen \etal \cite{fixsen} result may be approximated by $T_{0} \approx 2.7\,{\rm K}$, corresponding to a CMB photon number density $n_{\gamma0} \approx 400\,{\rm cm}^{-3}$ (compared to the more accurate results, $T_{0} = 2.7255\,{\rm K}$ and $n_{\gamma0} \approx 411\,{\rm cm}^{-3}$).

\subsection{Effect on BBN}

What happens to BBN when we allow for extreme variations in $\eta_B$?
As noted earlier, the most important effect of increasing
$\eta_B$ is to increase the primordial $^4$He mass fraction at the expense of hydrogen.  One might imagine that a universe
in which stellar evolution begins with almost pure $^4$He might be less hospitable to life. 
For example, Hall et al. \cite{Hall} pointed out that in such a universe, halo cooling takes longer, stellar lifetimes are reduced, and there is less
hydrogen to support organic chemistry.  (The calculations in Ref. \cite{Hall} are focused on
variations in the weak scale, rather than the magnitude of the baryon asymmetry).  However, even extreme increases in $\eta_B$
do not produce primordial $^4$He mass fractions close to 100\%.  For example,
a value of $\eta_B$ as large as $10^{-3}$ (more than six orders of magnitude larger than
the observed value) yields a $^4$He mass fraction of only 0.4 \cite{Scherrer}.

Large values of $\eta_B$ also open up the possibility of producing heavier elements in BBN.  Consider first the CNO elements.
In standard BBN, these are produced in very small amounts, with abundances relative to hydrogen of CNO/H $\sim 10^{-15} - 10^{-14}$ \cite{Coc}.
However, the abundances of these elements are an increasing function of $\eta_B$, peaking at CNO/H $\sim 10^{-8}$ for $\eta_B \sim
10^{-5}$,
and decreasing for larger values of $\eta_B$ \cite{wfh}.  Even a small primordial abundance of  CNO/H could affect the evolution of the first generation of stars,
as noted in Ref. \cite{CC}; this evolution begins to change when CNO/H increases above $10^{-11}$.  Nonetheless, it seems unlikely that
such a change would affect the ability of the Universe to harbor life.  For $\eta_B > 10^{-5}$, the abundance of the CNO elements begins
to decrease, as the nuclei are converted into even heavier elements \cite{wfh,Scherrer}.  However, even extreme increases in the value
of $\eta_B$ result in only trace amounts of such heavy elements.  In terms of models that can support life, it does not appear that
BBN provides a useful upper bound on $\eta_B$, and certainly not a bound competitive with arguments from structure/galaxy/star
formation.

Now consider BBN in the limit of very low values for $\eta_B$.  In this limit, the $^4$He abundance becomes negligible, while
$^2$H increases, reaching a peak abundance of order ${\rm D/H} \sim 10^{-2}$ when $\eta_B \sim 2 \times 10^{-12}$.  For smaller values of
$\eta_B$, even the deuterium abundance decreases as $\eta_B$ is reduced, yielding, in the
limit $\eta_B \rightarrow 0$, a primordial Universe consisting essentially
of pure hydrogen.  The reduction in primordial helium for small values of $\eta_B$ is likely to reduce the cooling of galaxies that results from the collisional
excitation of ionized helium, but this is unlikely to have a major impact
\cite{Hall}.  On the other
hand, a significantly larger abundance of deuterium would lead to enhanced molecular cooling through an increase in the HD
abundance \cite{LeppShull}. 
While interesting, this is also unlikely to affect the prospects for a life-bearing universe.

Our conclusion then, is that BBN provides essentially {\it no} constraints
on Universes with different values of $\eta_B$.  The formation of stars
and planets and the development of life is nearly completely insensitive to
variations in the primordial element abundances, at least within the ranges
of $\eta_B$ that we have considered here.

\subsection{Effect on Large Scale Structure: the Linear Regime}

In the standard model for structure formation, small initial fluctuations
in the density are imprinted on the matter and radiation
by inflation or some other
process early in the evolution of the universe.  When the universe is
radiation-dominated, these fluctuations cannot grow inside of the horizon;
subhorizon fluctuations begin to grow once matter dominates the radiation.
If $\delta \rho/\rho$ represents the magnitude of the fluctuation in the matter density relative
to the mean matter density, then after matter domination begins,
$\delta \rho/\rho$
grows proportional to the scale factor $a$,
\begin{equation}
\label{linear}
\delta \rho/\rho \propto a.
\end{equation}
Eq. (\ref{linear}) applies only as long as $\delta \rho/\rho \ll 1$; in this
case the density fluctuations are said to be
in the {\it linear} regime.  Once $\delta \rho / \rho > 1$, the
Universe enters the nonlinear regime, and the analytic solution
given by Eq. (\ref{linear})
no longer applies.  Numerical simulations are necessary to evolve the density
field further forward in time.  In the nonlinear regime, the fluctuations in the matter density
grow
much more rapidly, and the dark mattter ultimately collapses into halos.

This process applies in a straightforward way only to dark matter,
which is collisionless.  The baryons evolve in a more complicated way.
At high temperatures ($T \gg 10^3$ K) the matter is ionized, and the cross section
for scattering off of photons is very high.  Thus, the baryons are frozen
to the radiation background and baryonic density perturbations cannot grow.
As the temperature drops, the electrons become bound to the protons and to the primordial
helium nuclei in a process known as recombination.\footnote{Note
that this term is a bit misleading, as the electrons and atomic nuclei were
never ``combined" to begin with.}  At this point, the density
perturbations in the baryons can begin to grow along with the dark matter perturbations.
A further complication is that in the nonlinear regime, the baryonic
matter, unlike the dark matter, is not pressureless and can also
radiate away energy in the form of photons.  Thus, at late times the baryons
evolve very differently than the dark matter.  The end result is that
the baryons ultimately bind into fairly compact disks or ellipsoids (galaxies),
fragment into stars, and form planets, while the dark matter remains in the
form of diffuse halos surrounding the galaxies.

In considering the effect of changing $\eta_B$, we must therefore consider
the change in two key parameters:  the redshift of equal matter and radiation,
and the redshift at which recombination occurs.  However, redshifts are defined
relative to the present day, so they are not particularly useful in determining
whether a modified universe can support life, as we are not restricting life to form
at redshift zero as it does in our Universe.  Instead, we should examine the temperature
of equal matter and radiation and the temperature of recombination.
In our Universe, the temperature of equal matter and radiation, $T_{eq}$ is given in
terms of the present-day temperature, $T_0$, by
$T_{eq} = T_0(\rho_M/\rho_\gamma)_0$.  For the parameter values given at the beginning of
this section, we obtain $T_{eq} = 9000$ K.  How does this change when $\eta_B$ is altered?
To determine this, note that the redshift of equal matter and radiation is given
by this ratio of present-day densities:
\begin{equation}
1+z_{eq} = \left(\frac{\rho_{DM} + \rho_B}{\rho_\gamma + \rho_\nu}\right)_0.
\end{equation}
Here we are ignoring the fact that the neutrinos can become nonrelativistic at very late times.
Then we have:
\begin{equation}
1+\widetilde z_{eq} = \left(\frac{\rho_{DM} + F \rho_B}{\rho_\gamma + \rho_\nu}\right)_0.
\end{equation}

Using the values for the cosmological parameters above, we can trace out the effect of $F$
on $T_{eq}$.  We have $\rho_{DM}/\rho_B \approx 5$.  Thus,
$z_{eq}$ changes little for $F \lesssim 5$, while for $F \gtrsim 5$, we have $(1+\widetilde z_{eq}) = 
(F/5) (1+ z_{eq})$.  Then we have:
\begin{eqnarray}
\widetilde T_{eq} &\approx& T_{eq}~~~(F \lesssim 5),\\
\widetilde T_{eq} &\approx& \frac{F}{5} T_{eq}~~~(F \gtrsim 5).
\end{eqnarray}

Now consider the effect of altering $\eta_B$ on the recombination temperature $T_{rec}$.  While
recombination is a gradual process and does not occur suddenly at a single temperature, for
the purposes of this study it will be sufficient to take $T_{rec} \approx 3000$ K.  The
process of recombination depends primarily on the ratio of the photon temperature to the binding
energy of hydrogen, but there is also a residual dependence on $\eta_B$.  This dependence comes about
because $\eta_B^{-1}$ determines the number of photons per hydrogen atom; an increase in this number makes it
easier for photons to ionize the hydrogen, delaying recombination, while the reverse is true if the
number of photons per hydrogen atom decreases.  However, the temperature at which a given
ionization fraction is reached varies roughly logarithmically with $\eta_B$.  
This is a much
smaller effect than the change in $T_{eq}$ with $\eta_B$, so we will ignore it in what follows and take
the recombination temperature to be roughly insensitive to changes in $\eta_B$.

Now we can investigate the effect of changing $\eta_B$ on large-scale structure in the linear regime.
We will not consider
any possible changes in the magnitude of the primordial density fluctuations; we will assume
that these are unaltered.
We see that neither
of the parameters affecting large-scale structure are modified if $F \ll 1$, so the process of
structure formation, at least in the linear regime, proceeds in the same way as in our Universe.
The density of baryons relative to dark matter will be much lower, leading to fewer galaxies per
dark matter halo, but this by itself does not seem to be a barrier to the formation of stars and
planets.  In the opposite limit ($F \gg 1$), the universe will be become matter dominated early on,
but baryonic structure formation will not occur until the temperature drops down to $T_{rec}$, which is
essentially unchanged from its current value.  So in this case, too, we expect little change
to the process of structure formation.

\subsection{Effect on Large-Scale Structure:  the Nonlinear Regime}

Linear perturbation growth allows density perturbations to grow until $\delta \rho/\rho \sim 1$, but
it is the subsequent nonlinear perturbation growth that directly produces galaxies, stars, and planets.
Unfortunately, nonlinear perturbation growth is more difficult to characterize for two reasons.
First, it cannot be solved analytically and requires quite detailed numerical simulations.  Second,
nonlinear baryonic physics is quite a bit more complex than the behavior of collisionless
dark matter and can be difficult to simulate, even numerically.  In the absence of large-scale computer
simulations of alternate universes with different values of $\eta_B$, the limits discussed here should
be treated with some skepticism.

Tegmark et al. \cite{Tegmark} have examined systematically the effects on structure
formation of altering the baryon/dark matter density ratio, which, by assumption, is the same
as the change in the baryon/photon ratio. 
Consider first the lower bound on $\Omega_B/\Omega_M$.
Tegmark et al. argued that one can derive a lower bound based on the requirement that the
collapsing baryon disks be able to fragment and form stars.  If
the baryon to dark matter ratio becomes too small, then the baryonic matter is insufficiently
self-gravitating to allow fragmentation to
occur.  The limit derived in Ref. \cite{Tegmark} is $\Omega_B/\Omega_M \ga 1/300$,
which corresponds to the lower bound $\widetilde \eta_B > 1 \times 10^{-11}$.

In the absence of detailed numerical simulations, this bound should be treated with caution.  More
conservative lower bounds on $\eta_B$ were derived by Rahvar \cite{Rahvar}.  Star formation is significantly
suppressed at very low $\eta_B$ simply because there are not enough baryons around to form stars.  The
requirement that at least one star forms per galactic-sized halo mass gives $\widetilde \eta_B > 10^{-22}$.  One
can be even more conservative and require at least one star in the observable universe; this requires
$\widetilde \eta_B > 10^{-34}$ \cite{Rahvar}.

Tegmark et al. also derived an upper bound on $\widetilde \eta_B$ from Silk damping (also called diffusion damping).
Silk damping arises near the epoch of recombination from the diffusion of photons 
out of overdense (hotter) regions near the epoch of recombination.  As the photons diffuse, they scatter off of charged particles
and drag the baryons along with them, which tends to erase the baryonic density perturbations.  Tegmark et al. argue
that if the dark matter density were lower the baryon density at recombination, Silk damping would tend to erase all fluctuations on galaxy-sized
scales.  Thus, they derive the limit \cite{Tegmark} $\Omega_B/\Omega_{DM} \la 1$, corresponding to
$\eta_B < 3 \times 10^{-9}$.  Again, this limit should be treated with some caution; before the
discovery of dark matter, cosmologists did not consider purely baryonic models to be ruled out by an absence
of structure formation!

In summary, our results in this section do not point toward significant fine tuning of the baryon asymmetry parameter,
$\eta_B$.  Element production in the early universe provides essentially no limits on changes to $\eta_B$
from the point of view of the habitability of the Universe, while limits from structure formation are
either very weak, very speculative, or both.

\section{Summary and Conclusions}
\label{sec:conclusions}

For a dimensionless physical parameter such as the baryon asymmetry parameter, \etab, that could take on any value from $-\infty$ to
$+\infty$ (or, allowing for a swap in the definition of matter and  antimatter, from $0$ to $\infty$), zero might seem the most natural
choice.  However, the value of $\etab = 0$ corresponds to a symmetric Universe, a Universe with equal amounts of matter (baryons)
and antimatter (antibaryons), which is inconsistent with what we actually observe.  An overview of the problem was provided in
\S\ref{sec:intro}, where $\eta_B$ was defined and its relation to the baryon to entropy ratio was
discussed.  To address the question of whether a nonzero value for \etab~is, or is not, fine tuned, some ground rules are required.  These
were outlined in \S\ref{sec:groundrules}.  We evaluate fine-tuning in terms of the
ability of the Universe to produce stars, planets, and, ultimately, life.
As reviewed in \S\ref{sec:symmetric}, our Universe cannot be symmetric; observations strongly
indicate that $\etab \neq 0$.

An overview of the models that have been proposed
to account for $\etab \neq 0$ was offered in \S\ref{sec:sakharovetal}.  The variety of models in the literature suggests that virtually
any value of \etab, including the other ``natural" value of $\etab \approx O(1)$, could be ``predicted."  The observations most sensitive to \etab~are the abundances of the elements produced during BBN.  The dependence
of BBN on \etab~was reviewed in \S\ref{sec:bbn}, revealing that while the precise abundances vary significantly with \etab, over a very large range in \etab~only hydrogen and helium (\4he) emerge from the early evolution of the Universe with significant abundances.
The connection between \etab~and a variety of other cosmological parameters
was discussed in \S\ref{sec:cosmo}, and the effect of changing $\eta_B$ on the evolution
of the Universe
was examined in \S\ref{sec:alternate}.
While large changes in $\eta_B$ affect both primordial element production and the formation of galaxies and stars, it is only the
latter that allows us to suggest limits on the allowed range for $\eta_B$.
Our results indicate that universes with values of the baryon asymmetry parameter
that differ significantly from our own can form galaxies and stars (whose evolution can produce the heavy elements necessary for life), and
planets, capable of hosting life.
Thus, the value of $\eta_B$ can be varied by many orders of magnitude
without strongly affecting the habitability of the Universe,
a result that is ${\it not}$ suggestive
of fine-tuning.

It is likely that our Universe began with no baryon asymmetry (equal amounts of matter and antimatter), so that the initial baryon
asymmetry parameter had its ``natural" value of zero.  For a Universe like our own, conservation of baryon number, an exact symmetry
at very high temperatures, needed to be violated at some mass/energy scale in the very early Universe.  Processes such as those
described in \S\ref{sec:sakharovetal}, which must include baryon number nonconservation, resulted in the baryon asymmetry observed in
our Universe and in those alternate universes discussed here.  However, the baryon nonconservation required at high mass/energy
scales might also lead to nonzero (even if exponentially suppressed) baryon nonconservation at very late times in the evolution of the Universe.
If this were the case, then eventually, in a Universe that lives long enough, protons might decay (diamonds are
not forever!), so that the baryon number of the Universe (as well as the lepton number) would
revert back to its natural value of zero.  Ashes to ashes, dust to dust.\\

\begin{center}
{\bf Acknowledgments}\\
\end{center}

G.S. thanks K.M. Nollet, P.J.E. Peebles, and S. Raby 
for helpful and enlightening discussions.  R.J.S. thanks
J.F. Beacom for collecting and providing Gary Steigman's notes for this chapter.

\end{document}